# Self-Assembling Oxide Catalyst for Electrochemical Water Splitting

Daniel S. Bick[1,2], Andreas Kindsmüller[1,2], Deok-Yong Cho[3] , Ahmed Yousef Mohamed[3] ,
Thomas Bredow[4], Hendrik Laufen[1,2], Felix Gunkel[1,2], David N. Mueller[2,5], Theodor Schneller[1,2],
Rainer Waser[1,2,5] and Ilia Valov[1,2,5]*

[1]Institute for Materials in Electrical Engineering and Information Technology (IWE2), RWTH Aachen University of Technology, D-52074 Aachen, Germany

[2]JARA – Fundamentals of Future Information Technology, FZ Jülich, D-52425 Jülich, Germany

[3] IPIT & Department of Physics, Chonbuk National University, KOR-54896, Jeonju, Republic of Korea

[4]Mulliken Center for Theoretical Chemistry, Institute for Physical and Theoretical Chemistry, University Bonn, D-53115 Bonn, Germany

[5]Peter Grünberg Institute, FZ Jülich, D-52425 Jülich, Germany

**Summary**

Renewable energy conversion and storage, and greenhouse gas emission-free technologies are within the primary tasks and challenges for the society. Hydrogen fuel, produced by alkaline water electrolysis is fulfilling all these demands, however the technology is economically feeble, limited by the slow rate of oxygen evolution reaction. Complex metal oxides were suggested to overcome this problem being low-cost efficient catalysts. However, the insufficient long-term stability, degradation of structure and electrocatalytic activity are restricting their utilization. Here we report on a new perovskite-based self-assembling material $BaCo_{0.98}Ti_{0.02}O_{3-\delta}:Co_3O_4$ with superior performance, showing outstanding properties compared to current state-of-the-art materials without degeneration of its properties even at 353 K. By chemical and structural analysis the degradation mechanism was identified and modified by selective doping. Short-range order and chemical composition rather than long-range order are factors determining the outstanding performance. The derived general design rules can be used for further development of oxide-based electrocatalytic materials.

The demand on efficient and economically reasonable conversion and storage of energy from renewable power sources is an essential challenge and existential task for the modern society [1-3]. Emerging technologies like metal-air batteries [4-6], fuel-cells [4, 7-9] and electrolysers [10], which are mostly limited by the oxygen electrocatalysis, depend on reliable catalyst materials suited for long term application in alkaline environments. Candidates for precious metal free catalyst materials range from metal alloys [11, 12] to oxides [13-20] and nitrides [21]. Several perovskites have been suggested as economically reasonable catalysts for the oxygen evolution reaction (OER) and showed eligible overpotentials [16-17, 22-24]. Among these, the perovskite system $BaCoO_3$ (BCO) [25], and especially the double perovskite $Pr_xBa_{1-x}CoO_{3-\delta}$ (PBCO) has shown superior properties (low overpotential) and has been identified as one of the most promising OER materials [23, 26].

Application of oxides as OER electrocatalysts is limited by two general problems. The first one is their insufficient electronic conductivity. The use of conducting binder-materials is widespread to overcome this problem [23, 27-28]. However, this circumstance limits the application and complicates the evaluation of the chemical stability and the intrinsic properties of these electrodes, thus urging the development of binder free electrocatalysts. The choice of perovskites (for binder free catalyst) is therefore limited to highly electronically conducting material systems.

The second and main problem for all OER catalysts is the degradation of the electrocatalytic properties and loss of mass, related to chemical (leaching) and/or structural (amorphisation) changes during long-term operation. Despite several papers report on these issues, no study was dedicated to detailed understandings on degradation mechanism(s), structural and/or chemical changes and the factors influencing it. Moreover, most of these studies consider properties and performance measured at room temperature where the degradation processes (thermally activated) are kinetically hindered, whereas real operating temperature is supposed to be ~ 353 K.

The ideal material should have a high electrocatalytic activity, sufficiently high electronic conductivity (to prevent Ohmic loss and use of binders), and most importantly long time stability of structure/composition-related properties. In that sense the main challenge is to understand and prevent material destruction and especially loss of electrocatalytic properties occurring during water electrolysis.

The stability of perovskites in general has been studied mainly in terms of high temperature applications such as solid oxide fuel cell cathodes and membranes[29, 30]. It has been shown that segregation of Sr at high temperatures in some perovskites, i.e. LaSrCoO$_3$ and BaSrCoO$_3$ can be suppressed by doping with stable and high valent metal cations (V, Nb, Ti, Zr, Hf, Al) [31, 32].

At room temperature, loss of ferroelectric properties of perovskites has been reported and related to pinning of oxygen vacancies[33]. However, to date there are almost no studies on the degradation mechanism in perovskites during water electrolysis. Only recent works have shown, that the degradation of PBCO during oxygen evolution is related to its defect chemistry [34]. The degradation during OER is caused by chemical leaching of cations which leads as a consequence to amorphization of the catalyst's structure.

Here we show a new perovskite self-assembling material system BaCo$_{0.98}$Ti$_{0.02}$O$_{3-\delta}$:Co$_3$O$_4$, which exhibits higher current densities for the OER and over 10-fold increased lifetime in comparison to the most reliable electrocatalyst (PBCO) reported up-to-date. Importantly, all electrolysis experiments were performed at temperatures typical for industrial application i.e. 353 K. By systematic modification of chemical composition/doping and defect chemistry in binder free BCO perovskite-based catalyst films, simultaneous aging and characterisation with electrochemical methods, XRD, XPS, XANES and EXAFS we were able to identify the degradation mechanism related to cation leaching and amorphization. We demonstrate that the initial crystalline structure rapidly and completely transforms to an amorphous electrochemically active material which retains its electrochemical properties until its service life end.

The role and position of the cation dopant on the defect chemical structure of the BCO perovskite system is experimentally verified and supported by quantum chemical calculations. Its correlation to long-term stability of the OER catalyst is demonstrated. The self-assembled specific short-range order of the amorphous oxide stabilized during OER is responsible for the superior properties of the catalyst.

$BaCo_{0.98}Ti_{0.02}O_{3-\delta}:Co_3O_4$ was designed in accordance to our knowledge on the mechanism of perovskite degradation during OER in alkaline media. Based on the combination of analysis, calculation and aging experiments, a model is derived, which is able to explain the connection between short range order, defect chemistry and catalyst stability. This model is capable of making predictions to future OER catalyst material design.

**Thin film $BaCoO_3$-based electrocatalysts**

In our experiments we used as initial material BCO thin films with hexagonal structure and compared the effects of cation doping on A and/or B sites on the electrocatalytic activity and the long term stability. Thus, we compared the properties for BCO, $Pr_{0.2}Ba_{0.8}CoO_{3-\delta}$ (A-site dopant Pr) and $BaCo_{1-x}Ti_xO_{3-\delta}$ (B-site dopant Ti) with x = 0.02 – 0.05. Simultaneous A and B doping results however in an unstable material with properties inferior to individual A or B-site doped material.

We found that excess of $Co_3O_4$ (cobalt black) has notable influence on the long term stability. Therefore also the properties of same perovskite composite materials were compared with and without addition of cobalt black (CB). From all tested samples, the best performance for the oxygen evolution reaction in terms of electrocatalytic activity and long term stability were obtained for $BaCo_{1-x}Ti_xO_{3-\delta}: Co_3O_4$ with x = 0.02. As discussed below, the superior properties were found to be determined by a unique structural transition of the material enabled by doping, and leading to a particular short-range order, that stabilizes the catalyst's surface.

**OER Catalytic Performance and Stability**

Catalytic performance, but especially stability under OER conditions are crucial for the design of reliable OER catalysts. Whereas the electrocatalytic activity is relatively easy to compare (by overpotential and current density), benchmarking of the long-time stability of OER catalysts has been a problem, since practical methods for short-term testing of the long-time stability are not well established yet. End of service life tests (ESLT) are suggested to enable these lifetime estimations [34]. All compositions were deposited on platinized Si as thin films with a thickness of 100 nm (Figure 1a, structural information in Extended Data Fig. 1, further details in methods part). This thickness ensures a reasonable time for the ELST tests and same initial geometrical electrode surface to compare the properties of the anode materials for water electrolysis.

Pure BCO is degenerating its catalytic performance continuously during the ESLT (Extended Data Fig. 2a.). A-type doping (PBCO) improves the current densities as well as the stability of the catalyst over time. As reported before, the best results for $Pr_xBa_{1-x}CoO_{3-\delta}$ are shown for the composition with x = 0.2 i.e. $Pr_{0.2}Ba_{0.8}CoO_{3-\delta}$ which is retaining its catalytic performance until complete dissolution[34].

B-type substitution by Ti-cations in BCO also improves the long-time stability and leads to an increase of the current densities. Increasing the amount of Ti-dopant starting from 2 at.% of up to 4 at.% leads to improvement of the sample characteristics. However, for 5 at.% a larger degeneration in current density is observed, rendering this catalyst unstable.

We found empirically that excess $Co_3O_4$ added to the films improves the performance by tuning the chemical equilibrium during degradation. The amount of the excess $Co_3O_4$ was adjusted by variation of Co to Ba ratio in the precursor solution: 1/1, 4/3 to 2/1.

Compared to $BaCoO_{3-\delta}$, $BaCoO_{3-\delta}:Co_3O_4$ (no Ti-doping) shows improved performance. However, its electrocatalytic activity drops significantly during the long term tests. Doping by only 2 at.% Ti on the B-site, an exceptional increase of lifetime and complex performance was observed. $BaCo_{1-x}Ti_xO_{3-\delta}:Co_3O_4$ (BCT:CB) showed the best characteristics, accounting for the combination of all important factors i.e. electric conductivity, current density, long-term electrocatalytic stability. A comparison to some of the best perovskite OER catalysts reported in the literature e.g. PBCO, BSCF and $Co_3O_4$, is provided in Figure 1. At room temperature the electrocatalytic activity of BCT:CB 2 is within the same range as PBCO and BSCF, showing similar Tafel slope (~60 mV/dec) and overpotentials. The crucial advantage is however, the performance at 80 °C (industrially relevant) where all other tested perovskite materials show fast etching rates (mass-loss) and rapid degeneration of their electrocatalytic performance within only 200-300 cycles. In Figure 1d is shown how fast the initially ($t$ = 0 s axes) superior electrocatalytic properties of BSCF change. After this initial period BCT:CB 2 demonstrates the best performance with over 40% higher current densities ($j$ = 68 mA/cm$^2$), compared to PBCO samples ($j$ = 48 mA/cm$^2$) and over 260% compared to BSCF ($j$ = 26 mA/cm$^2$). The lifetime end of BCT:CB with 2 at.% Ti doping (BCT:CB 2) is extended from 4 x10$^4$ s up to 38 x10$^4$ s for 100 nm catalyst material in accelerated test conditions which corresponds to 10 time higher long-term stability.

The calculated etching rate (mass-loss) is 83 µm over 10 years in accelerated conditions, which is undercutting MnO$_x$ (6.7 mm 10 years, at RT)[35] by far and is competitive to General Electric's patented Ni/stainless steel anodes (114 µm over 10 years, industry conditions)[36].

Thus, BCT:CB 2 was found to be a reliable and cost-efficient OER catalyst with superior lifetime and electrocatalytic performance.

**Ti-doping - lattice position and chemical changes**

The defect chemical equation of Ti substitution in BCO using Kroeger-Vink notation is:

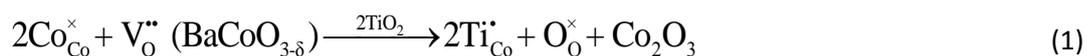

$$2Co_{Co}^{x} + V_{O}^{\bullet\bullet} \; (BaCoO_{3-\delta}) \xrightarrow{2TiO_2} 2Ti_{Co}^{\bullet} + O_{O}^{x} + Co_2O_3 \qquad (1)$$

Eq. 1 shows that substitution of Co$^{3+}$ by Ti$^{4+}$ on B-site will lead to lowering of the oxygen vacancy concentration. The substitution of Ti dopant on B-site (Co-site) is verified by XANES measurements at the Ti K-edge, identifying the perovskite atomic environment of the Ti dopant (Extended Data Fig. 3a.). Importantly, XPS Ti 2p spectra illustrate that Ti-dopant remains stable within the catalytic film and is not leaching into the solution during electrolysis/degradation (Extended Data Fig. 3b.).

The direct influence of Ti doping on the oxygen vacancy formation energy $E_d$ in general was calculated by DFT (Further details in SI section 1) as a model system for hexagonal BCO in the as-deposited state. For pure BCO, defect formation energies $E_d$ between 0.60 eV and 0.72 eV were obtained for $V_O^{\bullet\bullet}$ (Extended Data Table 1). In the vicinity of Ti-dopants, $E_d$ were calculated using three different oxygen sites with Ti-O distances of 1.87 Å (I), 4.61 Å (II) and 6.83 Å (III). The calculated $E_d$ decreases from 1.48 eV (I), 1.26 eV (II) to 1.23 eV (III). Thus, Ti-doping on B-site effectively increases the formation energy of the oxygen vacancies in the perovskite material, respectively decreases their number.

XPS measurements provide more detailed information about the chemical changes in BCO upon Ti-doping. The Ba 4d peak has binding energies (BE) of 89.8 eV in BaO [37] or 87.7 eV in $BaTiO_3$ [38]. Ti doping is changing the binding energy (BE) of Ba in both BCO (Extended Data Fig. 4a.1-5) and BCO:CB (Extended Data Fig. 4b.1-5). In our samples Ba 4d line is splitting in two BEs positioned at 88.3 eV (5/2, peak 1, A-site) and 89.0 eV (5/2, peak 2, A'-site). The peak at lower BEs represents Ba in an imperfect lattice environment (peak 1), and peak 2 at higher BE was interpreted as Ba in its perfect lattice environment[34]. Peak 1 to peak 2 area ratio is a measure for the imperfection of the lattice in respect to the ligand field around the Ba ion. Ti doping is inducing both a decrease in oxygen vacancy concentration (Eq. 1) and a distortion due to its larger size (0.0605 nm)[29] as compared to Co. These effects are competing in determining the perfection/imperfection of the lattice. The BE position of Ba 4d peak shifts by about 1 eV towards higher energy with ageing.

Co 3p peaks appear at the expected BEs of 60.2 eV for $Co^{2+}$ and 61.1 eV for $Co^{3+}$ [39] and do not shift upon doping and/or aging. The absence of chemical shift as opposed to the Ba case showcases the retention of short range order, i.e. where Co is kept in its octahedral environment. However, the ratio between $Co^{2+}$ and $Co^{3+}$ ions changes upon doping. In the initial, crystalline BCO $Co^{2+}/Co^{3+}$ is 1.6 and Ti-doping is diminishing this ratio to 0.97 in Ti-BCO. After ESLT aging (and related rapid and complete amorphisation) $Co^{2+}/Co^{3+}$ ratio becomes similar for both materials i.e. 0.46.

**Effect of excess $Co_3O_4$**

We found that excess $Co_3O_4$ is an important factor for the high electrocatalytic activity and stability of BCT:CB 2 (Co/Ba = 4/3). Both BCT:CB and BCT show constant electrochemical characteristics over time. However, BCT:CB 2 shows 100% higher current densities and 700% longer life-time compared to BCT 2 (Extended Data Fig. 2).

As for CB-free films, here Ba 4d shifts during aging by 1 eV towards higher BE in both CB-enriched material systems (Figure 2b1. to b2.). The chemical state of Ba for degraded samples (Figure 2b.2) is assigned to peak 2-component only. Accordingly, Ba positioned on imperfect lattice site (A'-site) is completely leached out. Most essentially, Ba dissolution is hindered in BCT:CB 2 at.%. Figure 2c. shows the remaining Ba content (calculation details in SI Section 2). While at the surface of BCO, BCT 2 and BCO:CB only 9.1 %, 14.5% and 19.7% of the original Ba content remains present, respectively, in BCT:CB 2 60.2 % of its original Ba content is retained upon amorphization. This significant difference in dissolution rate is another reason for the reliable electrochemical properties of BCT:CB 2.

As prepared BCT:CB shows again no shift in the Co 3p BEs, but we detect a higher fraction of $Co^{3+}$ due to the added $Co_3O_4$ with $Co^{2+}/Co^{3+}$ ratio = 0.83 (Figure 2a1.). However, in contrast to CB-free BCO and BCT, after aging the remaining $Co^{2+}$ fraction in CB-enriched materials was slightly increased, with a $Co^{2+}/Co^{3+}$ ratio of 0.5 (formally corresponding to $Co_3O_4$ stoichiometry) which appears to be a crucial factor for the material's stability (Figure 2a2.).

**Chemistry of Degradation**

The main problem of oxide electroactive materials is the loss of catalytic properties, related to their chemical and structural degradation. A main step in the degradation of perovskite materials during water electrolysis is their amorphization, triggered by A-site (e.g. Ba) leaching [34]. This process is thermally activated and therefore much faster at 80 °C. In the first seconds to minutes after OER reaction begins, the surface of the material undergoes a crystalline-amorphous transition. After less than 5% of the service life-time the complete layer transited to amorphous state. However, our BCO-based materials do keep a short-range order structure, despite the long-range structure was lost. Hence, the long-range structural transition alone is not necessarily destroying the electrocatalytic performance. The essential factor, in this respect, is the remaining short-range order and the related stability of the chemical composition in the amorphous film.

The hexagonal short-range structure of BCT:CB 2 transits during electrolysis to tetragonal tungsten bronze type one and remains stable until life-time end. Referring to Johnsson and Lemmens [40], partial occupation of the A-site in $ABO_3$ perovskite structures can result in the formation of tungsten bronzes, which is understood as a transition between perovskites and ilmenites. The transition from a perovskite to a tungsten bronze structure includes a rotation of an octahedron, which is enabled by the removal of an A-site cation and a distinct status of defect concentration. This rotation is forming an additional tetrahedron site. [41] As shown in Figure 3a, the result of such a mechanism for BCT:CB 2 was evidenced by EXAFS experiments, which allows to study element-specific short-range order even in amorphous materials. We studied the short range ordering effect with ongoing degradation by identifying the coordination shells around Co cations. For as prepared samples the radial distribution functions (RDF) have a line shape close to the one reported in the literature for $Pr/LaCoO_3$ [42] (different interatomic distances due to cubic structure/corner sharing octahedra in reference) and show a direct influence of Ti doping in both BCO:CB/BCT:CB material systems on the first coordination shell (CS). The 1st CS is built by the oxygen octahedron. Doping with 2 at% Ti on the B-site directly broadens the feature of the Co-O bond at 1.4 Å reduced interatomic distance and illustrates the disorder (various bonding lengths) which is brought in the octahedral symmetry in the as prepared state. However, the octahedral symmetry signal is enhanced during aging tests, which implies further oxidation of Co, accompanied by a change in oxygen stoichiometry. BCT:CB 2 shows the highest magnitudes for the Co-O features and has coherently the highest order and, thus, lowest vacancy concentration in the oxygen coordination shell after the aging procedure. With a comparison to the literature for $Co_3O_4$ [43-45], the 2nd CS can be attributed to the closest Co-Co distances in the network of octahedral (2.5 Å reduced distance) or tetrahedral (2.9 Å reduced distance) oxygen coordination. This shows that the hexagonal packing of the initial BCO phase is preserved locally, where Co-Co distances are the closest due to face sharing of the $CoO_6$ octahedra, whereas the formation of edge sharing octahedra creates tetrahedral sites, where the Co-Co distance is increased. Direct comparison between BCO:CB and BCT:CB 2 shows that BCT:CB 2 is forming significantly more tetrahedral sites hosting $Co^{2+}$ during degradation. XANES measurements have also confirmed the formation of tetrahedral $Co^{2+}$ (Extended Data Fig. 5). Though, the formation of tetrahedral sites is not caused by a transition to $Co_3O_4$ phase because the 3$^{rd}$ CS for $Co_3O_4$ must have a feature at 4.5 Å reduced distance due to the inherent alternating structure, where a tetrahedron shares an edge with the next octahedron. This feature is not enhanced during aging of BCT:CB 2.

Thus, the decomposition mechanism of perovskite-based OER electrocatalysts proceeds with formation of a short range structure related to tetragonal tungsten bronzes (P4/mbm) which is initiated by A-site leaching (Figure 3b.). BCT:CB 2 is able to stabilize this phase by providing the required oxygen stoichiometry, which means a stabilisation of the octahedral sites by pushing the coordination number towards 6, and at the same time setting the preconditions for octahedron rotation by providing existing tetrahedral sites occupied by $Co^{2+}$. In general, the rotation of octahedrons in a hexagonal perovskite, turning from face to edge sharing octahedral, degenerates A'-sites in defect lattice environment into A'-sites corresponding to Figure 3b2. Also, the rotation will generate tetrahedral sites by eg. two edge sharing octahedra (called C-site in Figure 3b2.). In principle, cubic perovskites are also able to generate tetrahedral sites via octahedron rotation.

BCT:CB 2 is capable to form this defect structure with a high ratio of Ba on stable, stoichiometric and 12-fold coordinated Ba sites, stoichiometric $CoO_6$ octahedrons and a high amount of Co bound in tetrahedral sites, which hinders the destructive Ba leaching and enables a variety of Co oxidation states at the surface for oxygen electrocatalysis. The stability of this defect structure is dependent on oxygen vacancy concentration as well as on Co excess.

**Conclusions**

We reported on perovskite- based self-assembling material $BaCo_{0.98}Ti_{0.02}O_{3-\delta}:Co_3O_4$ with superior long-time stability and OER electrocatalytic activity for alkaline water electrolysis at application relevant temperatures of 353 K. A systematic doping approach combined with chemical and structural studies allowed to identify the crucial factors for self-assembling electrocatalysts design. The short-range structural transition supported by a specific concentration of point defects – oxygen vacancies -enables to retain chemical composition and electrocatalytic performance. Hence, only one particular oxygen stoichiometry shows a stabilizing effect without depressing the catalytic performance. Short-range structural changes proceeds via rotation of $CoO_6$ octahedrons, forming additional tetrahedral sites in the lattice, thus enabling hexagonal to tetragonal transition that prevents A-Site leaching, i.e. prevents the major degradation process in these materials. We found that most important factors for OER catalytic materials are the short-range order and chemical composition rather than the long range order and crystallinity. These advanced understandings on perovskite structure related surfaces can be used as fundamentals for future design of oxide electrocatalysts.

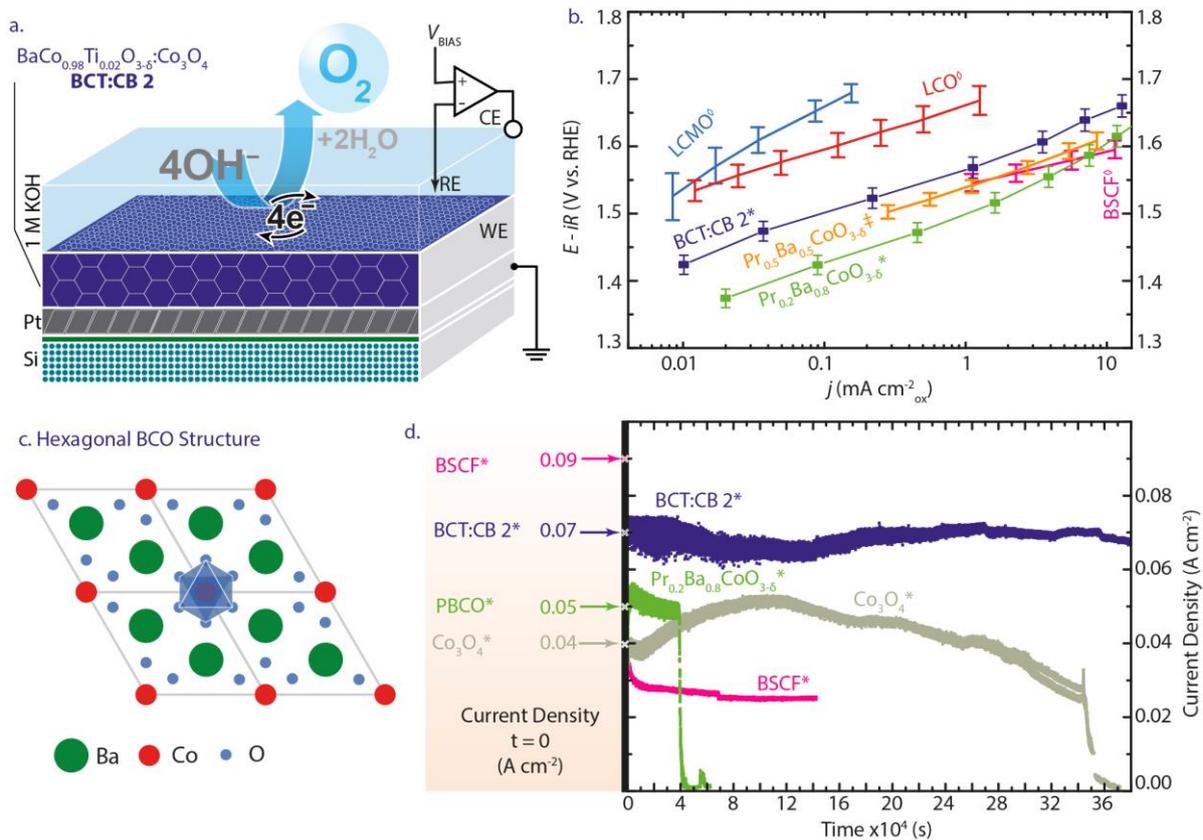

**Figure 1 | Chemical stability under OER conditions. a.** Schema of the sample and experimental setup. **b.** Steady state (Tafel) benchmark for several OER catalyst materials. Data and comparison is adapted from [17]. Data origins from: * for own work (1M KOH), ‡ for [23] (0.1M KOH), ◊ for [17] (1M KOH). All films are measured at RT. **c.** $Pr_{0.2}Ba_{0.8}CoO_{3-\delta}$ samples share the same hexagonal BCO structure. **d.** End of Service Life Test for BCT:CB 2 at.% Ti and $Pr_{0.2}Ba_{0.8}CoO_{3-\delta}$, polycrystalline and single phase thin films with 100 nm thickness on platinized Si substrates. BSCF and $Co_3O_4$ samples are not able to retain their catalytic activity until the service life end. ESLT is conducted at 353 K in 1M KOH.

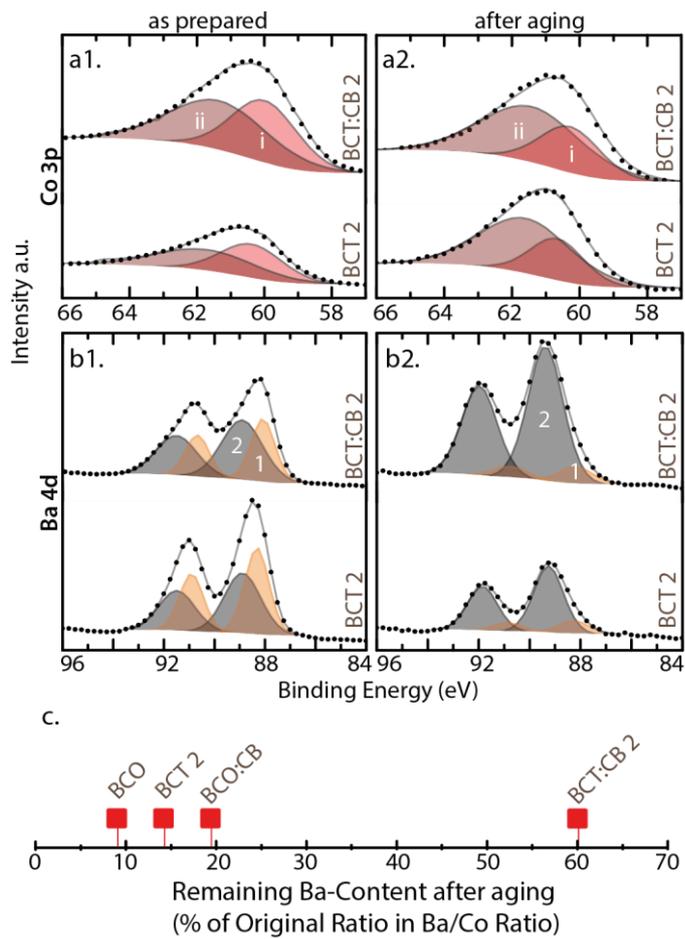

**Figure 2 | Reduced A-site leaching for BCT:CB.** Fitted XPS data for Co 3p spectra for BCT:CB and BCT thin films with 2 at.% Ti doping **a1.** as prepared and **a2**. after aging for 200 CV cycles up to 2V vs. RHE at 353 K in 1M KOH. Ba4d spectra **b1.** as prepared and **b2.** after aging. **c.** Remaining Ba content after aging in percent of original ratio in Ba/Co ratio.

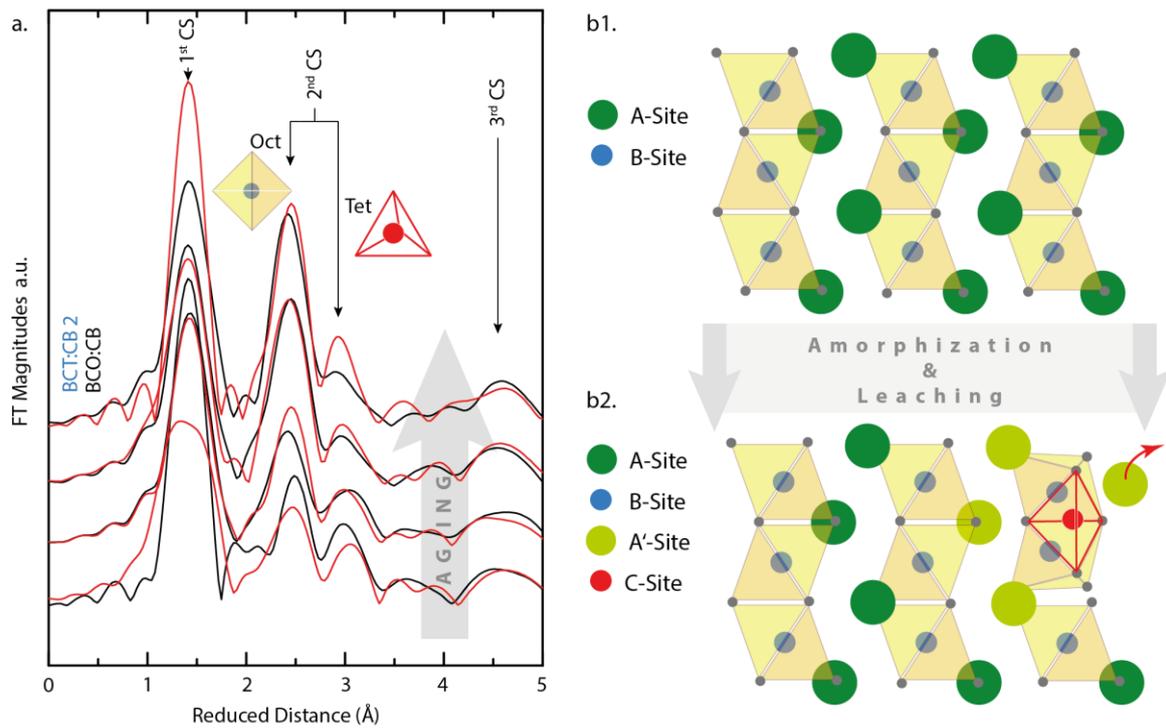

**Figure 3 | Degradation of perovskite structures during OER. a.** Fourier transformed EXAFS data during aging for BCO:CB (black) and BCT:CB 2 (red) for Co binding distances. **b.** Crystal structure draft of **b1.** a simple cubic $ABO_3$ perovskite and **b2.** a model for tetrahedral site forming by rotating $BO_6$ octahedrons.

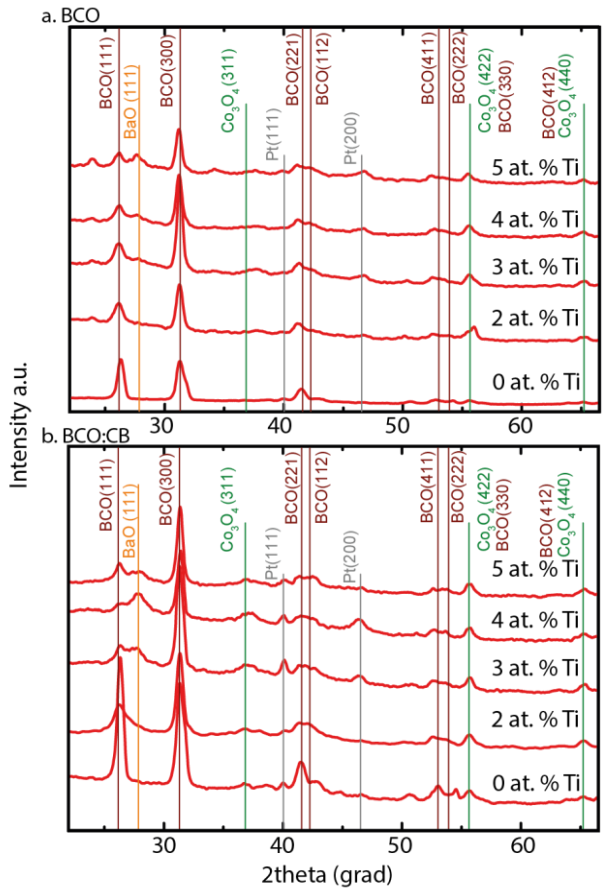

**Extended Data Figure 1 | Structural characterization.** Grazing incidence x-ray diffractograms of a. BCO and BCT 2-5 at.% Ti; b. BCO:CB and BCT:CB 2-5 at% Ti. Further Details in SI section 3.

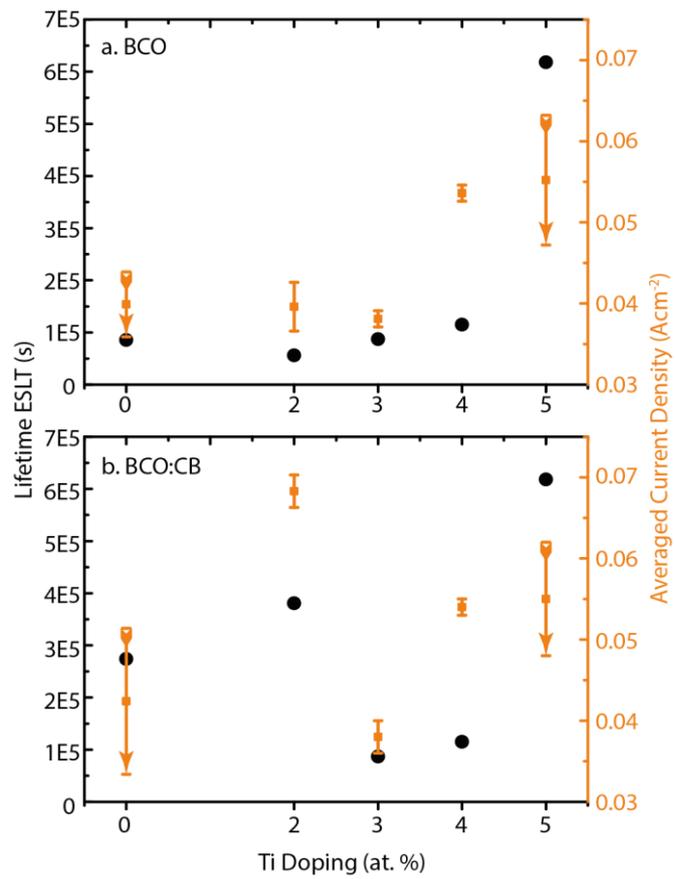

**Extended Data Figure 2 | Lifetime and Electrochemical Performance.** Results of End of Service Life Tests for a. BCO and b. BCO:CB showing the lifetime in seconds on y1 and the averaged current density at 2V vs. RHE, including its dispersion, on y2 over the Ti doping concentration. Further Details in SI section 4.

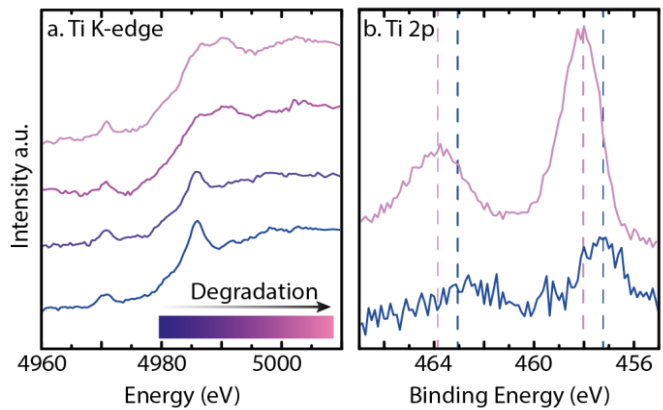

**Extended Data Figure 3 | Ti position before and after aging. a.** Ti K-edge of BCT:CB 2 thin films in different steps of aging. **b.** XPS data for Ti 2p spectra from BCT:CB 2 thin films before and after aging. Further Details in SI section 5.

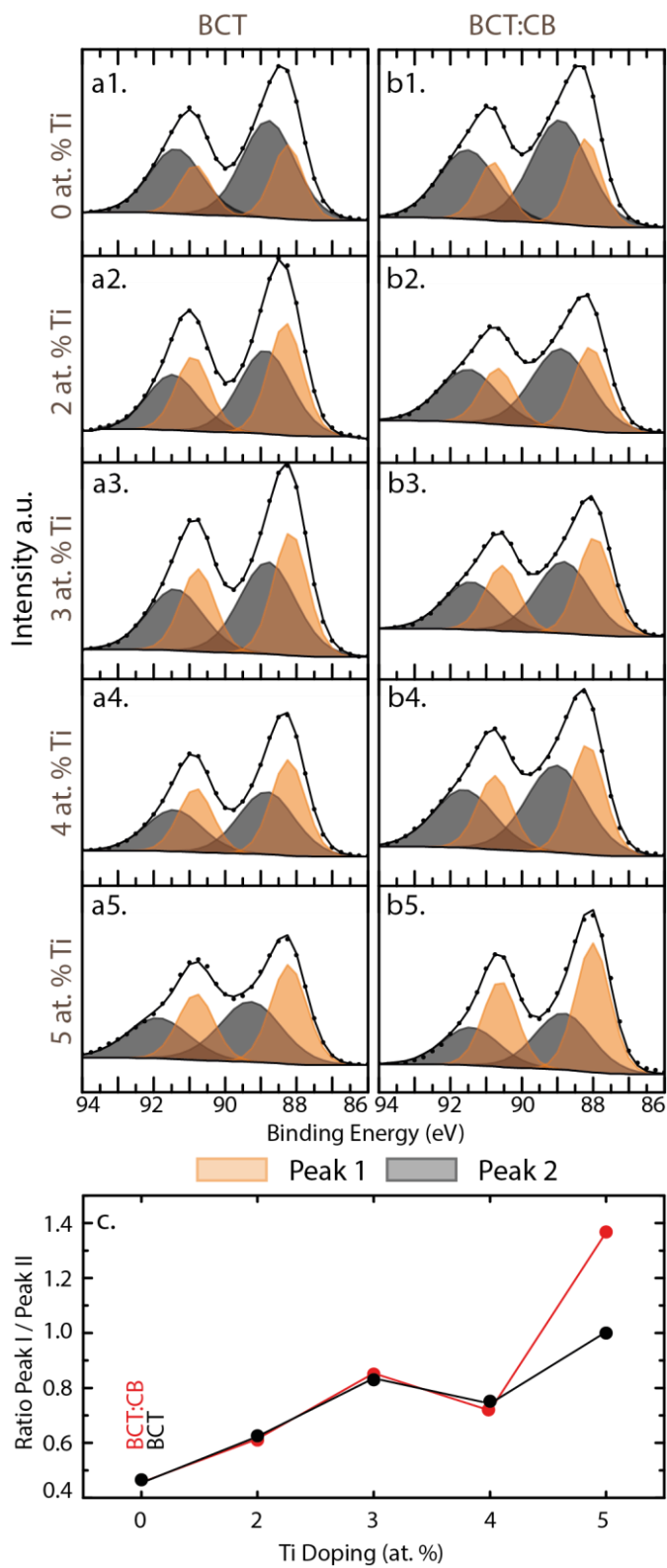

**Extended Data Figure 4 | Effect of Ti doping on the Ba binding energy.** Fitted XPS data for Ba 4d spectra at an angle of 75° from surface to detector on **a.** BCO; **b.** BCO:CB; **c.** Ratio of fitted Ba 4d components peak I and II of material systems BCT and BCT:CB. Further Details in SI section 6.

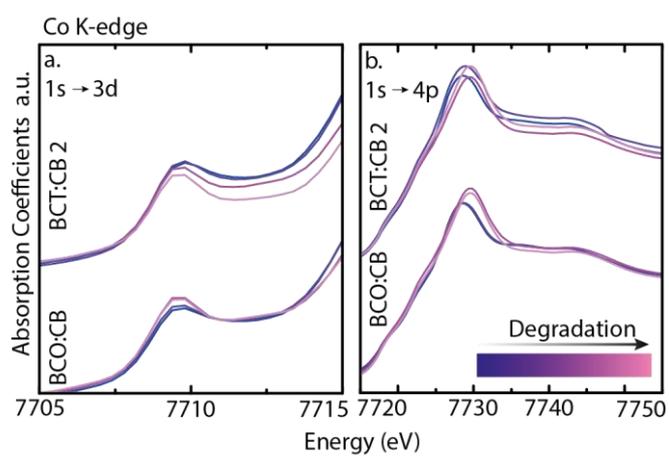

**Extended Data Figure 5 | Co oxidation state and Ti dopant incorporation.** Co K-edge XANES data sets during catalyst degradation for BCO:CB, BCT:CB 2 **a.** 1s to 3d transition; **b.** 1s to 4p transition. Further Details in SI section 7.

**Extended Data Table 1 | Defect Formation Energy.** Calculated O vacancy defect formation energies (eV) for $Ba_{16}Co_{16}O_{48}$ and $Ba_{16}Co_{15}TiO_{48}$ with a look on several O-Ti distances (Å) in the supercell.

| Supercell | $E_d$ (eV) | O-Ti distance (Å) |
|---|---|---|
| $Ba_{16}Co_{16}O_{48}$ | 0.60 | |
| $Ba_{16}Co_{15}TiO_{48}$ | 1.48 | 1.87 |
| $Ba_{16}Co_{15}TiO_{48}$ | 1.26 | 4.61 |
| $Ba_{16}Co_{15}TiO_{48}$ | 1.23 | 6.83 |

**Methods**

**Synthesis and Doping**

For substrate preparation, a (100) Si wafer with a native oxide surface is first coated with 5 nm $TiO_2$ (adhesion layer) and afterwards coated with 30 nm Pt via magnetron sputtering. For Ti sensitive XAS measurements, the adhesion layer was replaced by 5 nm $AlO_x$. The stack was annealed at 1073 K for 5 min before preparation of the catalytic film. After annealing, the Pt layer shows surface roughness rms < 1 nm (AFM measurement) and is (111) textured. The thin film stack including the Pt layer is crucial for lifetime measurements (ESLT-method), because Pt will tear from the adhesion layer on first contact with the electrolyte [34]. Substrates were pretreated with acetone and isopropanol followed by water rinsing before chemical solution deposition (CSD) [46].

Precursor solutions were prepared according to a modified propionic acid based routine originally used by Hasenkox et al. [47] to synthesize alkaline-earth doped lanthanum manganite thin films. The designated amounts of commercial available metal acetates ($Pr(OOCCH_3)_3$, $Ba(OOCCH_3)_2$ and $Co(OOCCH_3)_2$) were suspended in a mixture of propionic acid and propionic acid anhydride to remove the water of constitution of the metal acetates. Ti-doping was realized by adding Titanium(IV) oxide bis(2,4-pentandionate) into the solution. The suspensions were refluxed for 1 h and the metal salts dissolved. These solutions were restocked to a final concentration of 0.15 M (to B-site cation).

For deposition of the films a two-step annealing process was used. The precursor solutions were spun on the substrate surface with 3000 rpm for 30 seconds and pyrolised on a hotplate at 623 K for 2 minutes. The coating and pyrolysis procedure was repeated until the desired film thickness was achieved, followed by crystallization of the film at 1073 K for 5 minutes in air. BCT:CB 2 thin films were 100 nm thick, corresponding to a catalyst loading of around 620 ng/$cm^2$. Surface roughness of the catalyst film was under 10 nm rms.

**Materials characterisation**

XRD pattern were measured using an X'Pert Pro diffractometer (PANalytical) with Cu Kα radiation. In grazing incidence geometry, a glancing angle of 0.5° was used.

X-ray photoelectron spectroscopy (XPS) data was recorded using a Versa-Probe XPS tool (PHI electronics). The recorded spectra were aligned to the C 1s signal at binding energy of 284.6 eV. XPS data were fitted with XPS Casa commercial software. For the fit procedure we have used Gaussian/Lorentzian (GL30) functions after subtraction of the fitted background (Shirley) within the physical constrains regarding FWHM and peak intensity. The energy resolution of the electron detector was better than 0.2 eV.

Hard X-ray absorption spectroscopy (XAS) was performed at the 10C beamline of Pohang Light Source. The fluorescence yields of 100 nm-thick BCT:CB 2 and BCO:CB films were collected using a Si detector while the incident X-rays grazed the films with an angle of 10°. The X-ray absorption near-edge structures (XANES) were normalised by the edge jumps. Fourier transformation (FT) analyses of the extended X-ray absorption fine structures (EXAFS) were processed using the UWXAFS package [48]. The FT on the k2-weighted EXAFS oscillations (k: electron momentum) at Co K-edge, was done within a range of k = 3-13 Å-1 and the results are displayed in Fig. 3 as functions of reduced interatomic distance (before the scattering phase corrections).

**Electrochemical measurements**

All electrochemical measurements were performed in a membrane-free electrochemical glass-cell with three electrode geometry using a Pt counter electrode and Hg/HgO (in 1 M KOH) as reference electrode (Radiometer Analytical). In order to establish a corrosive environment, 1M KOH was used as electrolyte. The End of Service Life Test (ESLT) was performed with a scan rate of 100 mV/s between 1 V and 2.1 V vs. reversible hydrogen electrode (RHE) at 353 K. All data were iR corrected based on impedance measurements. More information about ESLTs has been published elsewhere [34]. ESLT Plots are showing the current density at 2 V vs. RHE over time. The sharp current decrease in the plot indicates the end of the service life. The test is associated with accelerated conditions compared to chronopotentiometric or -amperometric methods because of the periodic restart of the OER.

**Stepwise aging**

Sample aging was performed in steps. Aging steps were set to 50 cyclic voltammetry (CV) cycles at RT, 100 CV cycles at 353 K and 200 CV cycles at 353 K electrolyte (1 M KOH) temperature.

**Computational Method**

The quantum-chemical calculations were performed with the grid-based projector-augmented wave (GPAW) program [49-51]. The electronic wave function was described by a plane-wave basis set delimited by a relatively large cutoff energy of 900 eV. In order to account for effects of one-center electron exchange which are not well described by standard GGA functionals, the effective GGA+U correction by Dudarev [52] was applied to the PBE functional [53]. The value of parameter U was set to 3.0 eV in order to reproduce the electronic band gap obtained by a hybrid method, 1.2 eV. For this reference calculation the plane-wave program VASP version 5.4.1 [54] was employed. The same cutoff energy and the corresponding PAW implementation [55] were used. The PWXPW method was employed that provided accurate results for fundamental band gaps of transition metal compounds in a previous study [56]. Here it must be noted that hybrid calculations are feasible for the primitive unit cell but not for the supercell within the present limitations of computer resources.

The unit cell parameters of $BaCoO_3$ (spacegroup P63/mmc) were taken from Ref. [57] and kept fixed in the structure optimizations. Oxygen vacancies were created in a 2x2x2 supercell containing 16 formula units. For this supercell, sampling in reciprocal space was performed applying a Γ-centered 3x3x4 Monkhorst-Pack grid. This corresponds to a 6x6x8 grid for the primitive unit cell which proved to be sufficient to converge the total energy within 0.01 eV. In a previous TB-LMTO-ASA study Felser et al. reported on a ferromagnetic coupling within the quasi one-dimensional $Co^{4+}$ chains with low-spin d5 configuration [58]. This result was obtained within the spin-polarized local density approximation. With the present PBE+U method, the antiferromagnetic state with alternating spins along the Co chains is 0.07 eV per cell more stable than the ferromagnetic state. The spin density on the Co atoms is close to 1.0 in agreement with the low-spin state. It was checked that PBE without on-site correction the ferromagnetic state as ground state, and the initial electron configuration with antiparallel spins resulted in a diamagnetic state during the self-consistent field procedure.

**Founding Sources**

D.-Y. C. acknowledges the supports from Ministry of Science, ICT and Future Planning of Korea (NRF-2015R1C1A1A02037514).

**Author Contributions**

D.S.B. and I. V. created the idea, coordinated the experiments and wrote the manuscript. D.S.B. and H. L. performed the electrochemical measurements. A. K. conducted the XPS measurements. D.-Y. C. and A. Y. M. contributed XAS measurements and XANES / EXAFS analysis. T. B. performed the DFT calculations. R.W. and I.V. supervised the research. All authors contributed to discussion of the experimental results and improving the manuscript.

**References (Experimental)**

**Supplementary Information**

**Supplementary Methods**

Section 1 | DFT calculations:

The formation of oxygen vacancies in the undoped material system according to the Kröger-Vink notation in Equation S1 will result in the removal of an oxygen atom according to Equation S2.

$$O_O^\times + 2\,Co_{Co}^\times = V_O^{\bullet\bullet} + 2Co'_{Co} \qquad \text{Eq. (S1)}$$

$$Ba_{16}Co_{16}O_{48}(s) = Ba_{16}Co_{16}O_{47}(s) + \tfrac{1}{2}\,O2(g) \qquad \text{Eq. (S2)}$$

Since antiferromagnetic and ferromagnetic states were nearly degenerate in the non-defective BCO bulk, various possible spin configurations were tested for a $Ba_{16}Co_{16}O_{47}$ supercell. The total magnetization of the wave function was allowed to change from the initial configuration in the self-consistent field procedure and the structure optimizations. All atomic positions within the supercell were allowed to relax.

In the lowest-energy spin configuration (total magnetization 18) the spin density of two Co atoms increased from +1 to +2. This would correspond to a high-spin $s^0d^6$ configuration after a formal reduction from $Co^{4+}$ to $Co^{3+}$ by the two electrons left behind by the removed oxygen atom. A spin configuration where the two $Co^{3+}$ ions are antiferromagnetically coupled is only 0.05 eV less stable.

Ti was incorporated into BCO via Co substitution. This is experimentally supported and well justified because $Ti^{4+}$ ions are highly stable in oxides. It was further assumed that the lattice parameters of BCO are not affected by Ti substitution. The oxygen defect formation energy $E_d$ of BCT was calculated according to Equation S3 and will result in the removal of an oxygen atom according to Equation S4.

$$O_O^\times + Ti_{Co}^\times + 2\,Co_{Co}^\times = V_O^{\bullet\bullet} + Ti_{Co}^\times + 2Co'_{Co} \qquad \text{Eq. (S3)}$$

$$Ba_{16}TiCo_{15}O_{48}(s) = Ba_{16}TiCo_{15}O_{47}(s) + \tfrac{1}{2}\,O2\,(g) \qquad \text{Eq. (S4)}$$

Again various possible spin configurations were tested and full relaxation of all atoms in the supercell was performed. In all defect calculations the oxidation state of titanium remained 4+.

In order to investigate the long-range effect of Ti substitution on $E_d$, three different oxygen sites were selected with Ti-O distances of 1.87 Å (I), 4.61 Å (II) and 6.83 Å (III) according to Extended Data Table 1. Only the most stable spin configurations for each defect site are considered. The $Co^{3+}$ ions have spin densities close to 0, corresponding to a low-spin $s^0 d^6$ configuration in this case. Thus, much higher energies are required to form a oxygen vacancy in Ti-doped-BCO.

Section 2 | Calculation of Intensity Ratios in XPS

The remaining Ba-Content $r$ is given in % of the Ba/Co ratio in as prepared films. The Ba to Co ratio $a$ of as prepared films is calculated from the fraction of the peak Area of Ba $A^{Ba}_{as\ prep}$ to the peak area of Co $A^{Co}_{as\ prep}$ (Eq.S5). The Ba to Co ratio $b$ of as prepared films is calculated from the fraction of the peak Area of Ba $A^{Ba}_{aged}$ to the peak area of Co $A^{Co}_{aged}$ (Eq.S6). The fraction of $b$ and $a$ times 100 is $r$ in % (Eq.S7).

$$\frac{A^{Ba}_{as\ prep}}{A^{Co}_{as\ prep}} = a \quad (Eq.S5)$$

$$\frac{A^{Ba}_{aged}}{A^{Co}_{aged}} = b \quad (Eq.S6)$$

$$\frac{b}{a} \cdot 100 = r \quad (Eq.S7)$$

Section 3 | XRD:

BCT:CB contains a hexagonal perovskite phase BCO (PDF 00-048-1773) and a cubic spinel phase $Co_3O_4$ (PDF 00-042-1467). Extended Data Fig. 1 shows XRD data for BCO and BCO:CB structured polycrystalline material systems (structure along (111) oriented Pt). Samples with 4 and 5 at.% Ti do show an additional phase of BaO. This is also observed in BCO:CB (from 3 at% Ti and above). The formation of spinel $Co_3O_4$ is explicitly indicated by the (311) signal.

Section 4 | ESLT Plot details:

Extended Data Figure 1 is pointing out the stability of related catalysts films by showing the overall lifetime during ESLT with 100 nm film thickness and the total averaged current density, including its dispersion, over its lifetime. The dispersion gives a measure, if the catalyst performance is constant over its lifetime, since a regression in catalytic performance will lead to a decrease of the current density at the considered potential. Long lifetimes must correlate with small dispersion and high current densities to ensure the achievement of a reliable OER catalyst.

**Supplementary Discussion**

Section 5 | XANES Ti K-edge analysis:

Ti K-edge lineshape of as prepared BCT:CB is very similar to CaTiO$_3$ in direct comparison to the literature [59]. The lineshape is changing to SrTiO$_3$-like after the 2$^{nd}$ aging step. Hence, Ti is incorporated on the B-site of the perovskite structure during preparation. The pre-edge feature at photon energies of 4972 eV is indicating a coordination number of 5 to 6 in the first coordination shell of Ti [60]. The slight increase of intensity of the pre-edge peak with ongoing aging suggests a higher 6-fold coordination of the first coordination shell. The transition from CaTiO$_3$-like to SrTiO$_3$-like is interpreted as a transition from a more imperfect/distorted (CaTiO$_3$) to a more perfect/undistored (SrTiO$_3$) state of the oxygen coordination in the perovskite structure. The observed change in XPS Ti 2p spectra which show shifts of 0.7 eV to higher BEs during aging in material systems with and without Co$_3$O$_4$ show a similar incorporation of the Ti-dopant in these composites, however signal to noise ratio is quite low at these small amounts of dopant. The Ti 2p signal intensity is increasing during aging due to a higher Ti fraction at the surface because of Ba leaching.

Section 6 | Ba 4d XPS spectra

Ti doping is changing the binding energy (BE) of Ba. Ba 4d spectra BCO (Extended Data Fig. 4a.1-5) and BCO:CB (Extended Data Fig. 4b.1-5) show a comparable change in the ratio of the fitted components peak 1 and 2. The spectra are fitted with a spin-orbit splitting of 2.6 eV. In both systems, the ratio of peak 1/peak 2 proceeds congruent up to 2 at% Ti (Extended Data Fig. 4c). Accordingly, both material systems react comparably on Ti doping.

Peak 1/peak 2 ratio increases for 2 at.% and 3 at.% doping but drops for 4 at.% Ti-doping. We interpret this decrease as indication for reaching the Ti solubility limit within BCO and reaching a level of maximum compensation of oxygen vacancies. Both material systems do form individual fractions of a Ba oxide phase at 5 at.% Ti doping, which is the reason for a split in 1/2 ratio.

Section 7: XANES measurements

XANES measurements at the Co K-edge (Extended Data Fig. 5) depict a sharpening of the Co pre-edge feature (1s to 3d transition) with ongoing degradation. This sharpening is more progressed for the BCT:CB 2 samples compared to the BCO:CB samples. The sharpening of this feature is due to the formation of tetrahedral $Co^{2+}$ because the tetrahedral ligand field is allowing dipole transitions in contrast to the octahedral symmetry, where only quadrupole transitions may take place, which are weaker in intensity [61]. The peak position of the main peaks (1s to 4p) near 7729 eV shown in Figure 3b. is shifting to higher photon energies by $\Delta = +1.1$ eV during degradation for all observed material systems. This is possibly a signature of a further oxidation of the Co cations.

**References (SI)**